\documentclass[aps,prc,showpacs,nofootinbib]{revtex4}

\usepackage{epsfig}
\usepackage{graphicx,amsmath}
\newcommand\ba{\begin{eqnarray}}
\newcommand\ea{\end{eqnarray}}
\newcommand\nn{\nonumber}
\newcommand{\be}{\begin{equation}}
\newcommand{\ee}{\end{equation}}

\newcommand{\bas}{\begin{eqnarray*}}
\newcommand{\eas}{\end{eqnarray*}}

\begin{document}
\title{Polarization effects in the reaction 
$e^++e^-\to \rho^+ +\rho^- $ and determination of the
$\rho -$ meson form factors in the time--like region}

\author{C. Adamu\v s\v c\'in}
%%%
\altaffiliation{Leave on absence from
\it Department of Theoretical Physics, IOP, Slovak Academy of Sciences, Bratislava, Slovakia}
%%%
\affiliation{\it DAPNIA/SPhN, CEA/Saclay, 91191 Gif-sur-Yvette Cedex, 
France}
%%%
\author{G. I. Gakh}
%%%
\altaffiliation{Permanent address:
\it NSC Kharkov Physical Technical Institute, 61108 Kharkov, Ukraine}
%%%
\affiliation{\it DAPNIA/SPhN, CEA/Saclay, 91191 Gif-sur-Yvette Cedex, 
France}
%%%
\author{E. Tomasi-Gustafsson}
%%%
\affiliation{\it DAPNIA/SPhN, CEA/Saclay, 91191 Gif-sur-Yvette Cedex, 
France}
%%%

\date{\today}

\vspace{0.5cm}
\begin{abstract}
The electron positron annihilation reaction into four pion production has been studied, through the channel  $e^++e^-\rightarrow \bar \rho+\rho $. The differential (and total) cross sections and various polarization observables for this reaction have been calculated in terms of the 
electromagnetic form factors of the corresponding $\gamma^*\rho\rho $ current. 
The elements of the spin--density matrix of the $\rho -$meson were also calculated. 
Numerical estimations have been done, with the help of 
phenomenological form factors obtained in the space--like region of the momentum transfer squared and analytically extended to the time-like region.
\end{abstract}

\maketitle
 
\section{Introduction}

The electromagnetic form factors (FFs) of hadrons and nuclei provide important information about the 
structure and internal dynamics of these systems. Although such measurements are carried out since many 
decades \cite{Ho62}, the hadron FFs are presently object of extended experimental studies, 
due, in particular, to the availability of high intensity (polarized) beams, polarized targets and hadron 
polarimeters which allow to reach very high precision and to access new kinematical regions. Polarization techniques have been applied and methods suggested long ago 
\cite{Re68} became feasible in the region of the momentum transfers where data help 
to discriminate between different theoretical predictions. The most surprising result recently obtained, concerns the electric FF of the proton, which turns out to be smaller than previously assumed, and shows a behavior with the momentum transfer squared, $Q^2=-q^2$, which differs from the magnetic FF \cite{Jo99}.

In case of deuteron, elastic FFs have been determined up to $Q^2$=1.9 GeV$^2$ in space-like (SL) region, (for a review, see for instance \cite{GG02}).  The individual determination of the three deuteron FFs requires the measurement of the differential cross section and at least one polarization observable, usually the tensor polarization, $t_{20}$, of the scattered deuteron in unpolarized $ed$ scattering. The data on the three deuteron FFs, charge $G_C$, quadrupole $G_Q$ and magnetic $G_M$ are better described by impulse approximation (including eventually corrections due to relativistic effects, meson exchange currents, $\Delta $ isobars..) and contradict QCD predictions, even at the largest $Q^2$ value experimentally reached, which corresponds to internal distances of the order of the nucleon dimension. The measurement of deuteron FFs in time-like (TL) region is beyond the present experimental limitations. Estimation of the cross section and various polarization observables for the reaction $e^++e^-\to d+\bar d$ was given in Ref. \cite{GTG06}, extending the model \cite{Ia73} to the deuteron. This model has proved work very well for the nucleon, both in SL and in TL regions. Vector meson dominance (VMD) models, in general,  are very successful in describing the hadron structure: they contain a small number of parameters, with clear physical meaning, and they can be extended to TL region, inducing the necessary imaginary component, through the resonance widths \cite{ETG05}.

Hadron electroweak properties have also been investigated in the framework of the constituent quark models based on the Hamiltonian light--front formalism \cite{Ca95}.

The determination of the pseudoscalar--meson FFs (for example, for pions and kaons) requires only cross section 
measurements. They have been extensively studied both in SL and TL regions (see, for example, \cite{BV}).

The light vector mesons are less known, because their experimental determination is more difficult, due to their short lifetimes. However, the $t$ dependence of the cross section for diffractive vector--meson 
electroproduction gives (model--dependent) information on the charge radius, and radiative decays such as 
$\rho^+\to \pi^+\pi^0\gamma$ allow to obtain their magnetic moment. The models based on the light--front formalism have been applied to the calculation of the electromagnetic 
properties of various hadrons, namely: the pion charge FF \cite{CJC}, the $\rho $--meson electromagnetic 
FFs \cite{BH92,K94,C95,MF97}, the vector and axial FFs of the nucleon \cite{CC91,YTW}, and the 
radiative, leptonic and semileptonic decays of both pseudoscalar and vector mesons \cite{J91}.

For a spin--one hadron, these models have a fundamental inadequacy: the rotational invariance of the 
electromagnetic current operator is not ensured by its one--body component alone (violation of the angular condition). This violation is quantified by the deviation of a quantity $\Delta (Q^2)$ from zero  \cite{C95}. The most studied spin--one hadron is the deuteron. The calculations of the electromagnetic FFs for  
spin--one hadron (deuteron and $\rho $--meson \cite{BH92,K94,C95}) showed that the quantity $\Delta (Q^2)$ 
is not zero and it increases with $Q^2$. But for the deuteron, which is a 
nonrelativistic system, it was found \cite{GKFF} that the effects of the violation of the angular condition  are small on FFs at the accessible values of $Q^2$. On the contrary, the $\rho $--meson is a 
relativistic bound system (the momentum of the constituent quark  is not small compared to the $\rho $--meson mass) and the violation of the angular condition has large effects on 
the $\rho $--meson FFs \cite{K94}. It is therefore interesting to measure the electromagnetic FFs of 
relativistic spin--one hadrons, as the $\rho $--meson. For this aim, the most simple 
reaction  is  the annihilation of an electron--positron pair into a $\rho^+\rho^-$ pair. The experimental investigation of the electron--positron annihilation into hadrons (at low energies it 
is multipion final state) can also provide important information about the internal structure of the mesons which consisted of the light quarks, their interactions and the spectroscopy of their bound states. In the low--energy region,  these reactions are source of important information on other problems: the precise calculation of the strong interaction contribution to the anomalous magnetic 
moment of the muon, the check of predictions for the hadronic tau--lepton decay, etc.

In the past, the experimental study of the reaction $e^+e^-\to hadrons$ was limited to the measurement 
of the total cross sections only. But recently, the construction of large solid angle detectors which can operate at high luminosity colliders opened new possibilities  for the investigation of the reactions $e^+e^-\to multihadrons$ \cite{AKH99}. In the energy region $1\leq W\leq 2.5~ GeV$ ($W$ is the total energy of the colliding beams) the process of
four pion production is one of the dominant processes of the reaction $e^+e^-\to hadrons$. For the first time the process of four pion production in $e^+e^-$ annihilation was detected in Frascati 
\cite{B70} and somewhat later in Novosibirsk \cite{KU72}. The energy dependence of the processes $e^+e^-\to \pi^+\pi^-\pi^0\pi^0$ and  $e^+e^-\to \pi^+\pi^-\pi^+\pi^-$
was studied at the VEPP--2M collider (Novosibirsk) with the CMD--2 detector \cite{AKH99} in the energy range 
1.05--1.38 GeV.  The 
process of multihadron production at large energies was also investigated with the help of the BABAR detector at 
the PEP--II asymmetric electron--positron storage ring using the initial--state radiation \cite{Au05}. A summary of the hadronic cross section 
measurements performed with BABAR via radiative return is given in Ref. \cite{De06}. The analysis of the differential distributions showed that  the $a_1(1260)\pi $ 
and $\omega\pi $ intermediate states dominate and that the relative fraction of the $a_1(1260)\pi $ state increases with the beam energy. 

From the theoretical point of view, the processes of meson production in the 
electron--positron annihilation were considered in a number of papers. Using the vector--dominance model, the authors of 
Ref. \cite{KUW71} investigated the reaction $e^+e^-\to mesons$ assuming two--body (or quasi--two--body) final states. 
In particular, they considered the $a_1(1260)\pi $ and $\rho^+\rho^-$ two--body final states. The estimation 
of the hadronic cross section was made under the assumption of unit FFs. It was emphasized that the 
comparison of the VMD model with experiment at low energies would not be conclusive on the 
basis of the magnitude and energy dependence of the cross section only. The  observation of $\rho -$meson pairs with larger cross sections than predicted in the paper would probably 
signify, according to the authors, the presence of appreciable magnetic and/or quadrupole couplings. Estimations of the cross sections of 
the processes $e^+e^-\to 3\pi, 4\pi$ were obtained using a VMD model, in Ref. \cite{AK72}. 
Due to the conservation of vector current the cross section of the $e^+e^-\to 4\pi$ process can be related 
to the probability of the $\tau\to 4\pi\nu_{\tau}$ decay. Therefore, all realistic models describing the first 
process, should also be applicable to the description of the latter one.  The 
free--parameter investigation of the branching ratios and distribution functions of the four decay modes of 
$\tau\to \rho\pi\pi\nu $, in terms of the effective chiral theory of mesons, agreed with the data \cite{Li98} assuming that  $a_1$ dominates in these four decay modes of the tau--lepton. We will consider the  $a_1$ production in $e^+e^-$ annihilation in a forthcoming work. In this paper we consider the reaction 
\be
e^++e^-\to \rho^++\rho^-.
\label{eq:eq1}
\ee
Following a model independent formalism developed for spin 1 particles in \cite{GTG06}, we calculate the differential (and total) cross sections and various polarization observables in terms of the electromagnetic FFs of the corresponding $\gamma^*\rho\rho $ current. The elements of the spin--density matrix of the $\rho -$meson are also calculated.

The estimation of various 
observables is done on the basis of a simple VMD parametrization for $\rho $--meson FFs. As no data exist, the parameters were adjusted in order to reproduce the existing theoretical predictions in SL region \cite{MF97} where the $\rho -$meson electromagnetic FFs were 
calculated, both in covariant and light--front formalisms with constituent quarks.   The parametrization was then analytically extended to the TL region. The experimental determination of $\rho$-meson FFs in TL region, although challenging due to the small counting rate, 
is in principle possible at electron positron rings, such as Frascati, Novosibirsk and Bejing.
%%%%%%%%%%%%%%%%%%%%%%%%%%%%
\section{Formalism}
%%%%%%%%%%%%%%%%%%%%%%%%%%%%
In the one-photon approximation, the differential cross section of the reaction (\ref{eq:eq1}) in terms of the leptonic 
$L_{\mu\nu}$ and hadronic $W_{\mu\nu}$ tensors contraction (in the Born approximation we can neglect the electron mass) 
is written as
\be
\displaystyle\frac{d\sigma}{d\Omega } = \displaystyle\frac{\alpha^2\beta}{4q^6}L_{\mu\nu}W_{\mu\nu},
\label{eq:eq2}
\ee
where $\alpha=1/137$ is the electromagnetic constant, 
$\beta =\sqrt{1-4M^2/q^2}$ is the $\rho $--meson velocity, M is the mass of the $\rho$-meson and $q$ is the four 
momentum of the virtual photon, $q=k_1+k_2=p_1+p_2$, (note that 
the cross section is not averaged over the spins of the initial beams).

The leptonic tensor (for the case of longitudinally polarized electron beam) is
\be
L_{\mu\nu}=-q^2g_{\mu\nu}+2(k_{1\mu}k_{2\nu}+k_{2\mu}k_{1\nu}) +
2i\lambda \varepsilon_{\mu\nu\sigma\rho}k_{1\sigma}k_{2\rho}\ , 
\label{eq:eq3}
\ee
where $\lambda$ is the degree of the electron beam polarization (further we assume that the electron beam is completely 
polarized and consequently $\lambda=1$). 

The hadronic tensor can be expressed via the electromagnetic current $J_{\mu}$, describing the transition 
$\gamma ^*\rightarrow \rho^-\rho^+ $ 
\be
W_{\mu\nu} = J_{\mu}J^*_{\nu}.
\label{eq:eq4}
\ee

The expression for the hadron tensor $W_{\mu\nu}$, in terms of the $\rho-$ meson electromagnetic FFs, is calculated using the explicit form of the electromagnetic current $J_{\mu}$. The spin--density matrix of a $\rho-$  meson is composed of three terms, corresponding to unpolarized, vector and tensor polarized $\rho-$ meson:
\be
U_{1\mu}U^*_{1\nu}=-\left(g_{\mu\nu}-\displaystyle\frac{p_{1\mu}
p_{1\nu}}{M^2}\right) +\displaystyle\frac{3i}{2M}
 \varepsilon_{\mu\nu\rho\sigma}s_{\rho} p_{1\sigma}
+3Q_{\mu\nu}.
\label{eq:eq5}
\ee 
Here $s_{\mu}$ and $Q_{\mu\nu}$ are the $\rho-$  meson polarization four vector and quadrupole tensor, respectively. The four vector of the $\rho-$  meson vector polarization $s_{\mu}$ and the $\rho-$ 
meson quadrupole--polarization tensor $Q_{\mu\nu }$ satisfy the following conditions:
 $$s^2=-1,~ sp_1=0,~Q_{\mu\nu}=Q_{\nu\mu}, \ \ Q_{\mu\mu}=0, ~
p_{1\mu}Q_{\mu\nu}=0\ .$$ 

We consider the case when the polarization of the second $\rho-$meson in the reaction (\ref{eq:eq1}) is not measured.

Taking into account Eqs. (\ref{eq:eq4}) and (\ref{eq:eq5}), the hadronic tensor in the general case can 
be written as the sum of three terms
\be\label{eq:eq9}
W_{\mu\nu}=W_{\mu\nu}(0)+W_{\mu\nu}(V)+W_{\mu\nu}(T),
\ee
where $W_{\mu\nu}(0)$ corresponds to the case of unpolarized particles in the final state and $W_{\mu\nu}(V)$ $(W_{\mu\nu}(T))$ 
corresponds to the case of the  vector (tensor) polarized $\rho-$  meson. 

%%%%%%%%%%%%%%%%%%%%%%%%%%%%%%%%%%
As the $\rho$--meson is a spin--one particle, its electromagnetic current is completely described by three FFs. Assuming P-- and C--invariance of the hadron electromagnetic interaction, this current can be written as \cite{AR77} 
\ba
J_{\mu}&=&(p_1-p_2)_{\mu}\left[ -G_1(q^2)U_1^*\cdot U_2^*+\displaystyle\frac{G_3(q^2)}{M^2}
(U_1^*\cdot q U_2^*\cdot q-\displaystyle\frac{q^2}{2}U_1^*\cdot U_2^*)\right ]
\nn \\
&&-G_2(q^2)(U_{1\mu}^*U_2^*\cdot q-
U_{2\mu}^*U_1^*\cdot q),
\label{eq:eq5a}
\ea
where $U_{1\mu}$ $(U_{2\mu})$ is the polarization four-vector describing the spin one $\rho^-$ 
($\rho^+$), and $G_i(q^2)$ $(i=1, 2, 3)$ are the $\rho $--meson electromagnetic FFs. The FFs $G_i(q^2)$ are complex functions 
of the variable $q^2$ in the region of the TL momentum transfer ($q^2>0$). They are related to the standard $\rho $--meson 
electromagnetic FFs:  $G_C$ (charge monopole), $G_M$ (magnetic dipole) and $G_Q$ (charge quadrupole) by
\be\label{eq:eq6}
G_M=-G_2, \ G_Q=G_1+G_2+2G_3, \ \
G_C=-\displaystyle\frac{2}{3}\tau (G_2-G_3)+ \left (1-\displaystyle\frac{2}{3}\tau 
\right )G_1, \
\ \tau=\displaystyle\frac{q^2}{4M^2}\ .
\ee

The standard FFs have the following normalizations:
\be\label{eq:eq7}
G_C(0)=1\ , \ \  G_M(0)=\mu_{\rho}\ , \ \ G_Q(0)=-M^2Q_{\rho}\ ,
\ee
where $\mu_{\rho}(Q_{\rho})$ is the $\rho $--meson magnetic (quadrupole) moment.

The explicit form of various contributions to the hadronic tensor has been derived in 
\cite{GTG06} for the deuteron case and can be applied here, replacing the deuteron mass and   FFs by the corresponding quantities for $\rho$ meson. 

The resulting expression for the unpolarized differential cross section in the reaction CMS is 
\be
\displaystyle\frac{d\sigma^{un}}{d\Omega }=\displaystyle\frac{\alpha^2\beta ^3}{4q^2}D, \
D=\tau (1+\cos^2\theta )|G_M|^2+\displaystyle\frac{3}{2}\sin^2\theta \left (|G_C|^2+
\displaystyle\frac{8}{9}\tau ^2|G_Q|^2 \right ),  
\label{eq:eq16}
\ee
where  $\theta $ is the angle between the momenta of the $\rho^- $--meson(${\vec p}$)
and the electron beam (${\vec k}$). Integrating this expression with respect to the $\rho $--meson angular
variables one obtains the following formula for the total cross section of the reaction (\ref{eq:eq1})
\be\label{eq:i1}
\sigma_{tot}(e^+e^-\to \rho^-\rho^+)=
\displaystyle\frac{\pi\alpha ^2\beta^3}{3q^2}\left [3|G_{C}|^2+4\tau
\left (|G_{M}|^2+\displaystyle\frac{2}{3}\tau |G_{Q}|^2\right )\right ]. \
\ee
As for the deuteron case, let us define an angular asymmetry, $R_{\sigma}$, with respect to the
differential cross section, $\sigma_{\pi/2}$, measured at $\theta =\pi /2$, 
\be
\displaystyle\frac{d\sigma^{un}}{d\Omega }=\sigma_{\pi/2}(1+R_{\sigma}cos^2\theta ),
\label{eq:i2}
\ee
where $R_{\sigma}$ can be expressed as a function of the $\rho $--meson FFs
\be
R_{\sigma}=\displaystyle\frac{2\tau \left (|G_{M}|^2-\displaystyle\frac{4}{3}\tau |G_{Q}|^2 \right )-3|G_{C}|^2}
{2\tau \left (|G_{M}|^2+\displaystyle\frac{4}{3}\tau |G_{Q}|^2 \right)+3|G_{C}|^2}.
\label{eq:eq20}
\ee
This observable enhances the difference between the terms containing the FFs, which have a sine squared and cosine squared dependence (\ref{eq:eq16}), therefore it is more sensitive to the different underlying assumptions on the $\rho $--meson FFs than the angular distribution itself. A precise measurement of
this quantity, which does not require polarized particles, would be
very interesting.

As in the SL region,  the measurement of the
angular distribution of the outgoing $\rho $--meson determines the modulus of the
magnetic form factor, but the separation of the charge and quadrupole form
factors requires the measurement of polarization observables \cite{ACG}. The
outgoing $\rho $--meson polarization can be determined by measuring the angular distribution of the $\rho^- $--meson decay products. 

As it was shown in Ref. \cite{DDR}, a nonzero phase difference between FFs of two baryons (with 
1/2 spins) leads to a non vanishing T--odd single--spin asymmetry normal to the scattering plane in the baryon--antibaryon 
production $e^+e^-\rightarrow B\bar B$. This is also valid for spin 1 hadrons.

To derive polarization observable it is necessary to define a particular reference frame. When considering the
polarization of the final particle, we choose a reference system  with the  $z$ axis along the
momentum of this particle (in our case it is ${\vec p}$). The $y$ axis is
normal to the reaction plane in the direction of ${\vec k}\times {\vec p}$; $x$, $y$ and $z$ form a 
right--handed coordinate system.

The cross section can be written, in the general case, as the sum of unpolarized and polarized terms, 
corresponding to the different polarization states and polarization directions of the incident and scattered particles:
\be\label{eq:eq21}
\displaystyle\frac{d\sigma}{d\Omega}=
\displaystyle\frac{d\sigma^{un}}{d\Omega}
\left [1+P_y+\lambda P_x+\lambda P_z+ P_{zz}R_{zz}+
P_{xz}R_{xz}+P_{xx}(R_{xx}-R_{yy}) +\lambda P_{yz}R_{yz}\right ],
\ee
where $P_i$, $P_{ij}$, and $R_{ij}$, $i,j=x,y,z$ are, respectively, the components of the polarization vector, tensor, and  of the quadrupole polarization tensor of the outgoing 
$\rho^- $--meson $Q_{\mu\nu}$, in its rest system and $\displaystyle d\sigma^{un}/d\Omega$ is the unpolarized differential  cross section. $\lambda$ is the degree of longitudinal polarization of the electron beam. It is explicitly indicated, in order to stress 
that these specific polarization observables are induced by the beam polarization.

Let us recall the expressions of the different polarization observables in terms of the $\rho $--meson FFs:

\begin{itemize}
\item  The vector polarization of the outgoing $\rho $--meson: 
\ba\label{eq:eq17}
P_y&=&-\displaystyle\frac{3}{2}\sqrt{\tau }\sin(2\theta )Im
\left [\left (G_C-\displaystyle\frac{\tau }{3}G_Q \right ) G_M^*\right ]/ D.
\nn \\
P_x&=&-3\displaystyle\frac{\sqrt{\tau }}{D}\sin\theta Re \left (G_C-\displaystyle\frac{\tau }{3}G_Q \right )G_M^*, \
P_z=\displaystyle\frac{3\tau }{2D}\cos\theta |G_M|^2.
\ea

\item The part of the differential cross section that depends on the tensor
polarization can be written as follows
\begin{eqnarray}
&\displaystyle\frac{d\sigma_T}{d\Omega}&=
\displaystyle\frac{d\sigma_{zz}}{d\Omega}R_{zz}+\displaystyle\frac{d\sigma_{xz}}{d\Omega}R_{xz}+
\displaystyle\frac{d\sigma_{xx}}{d\Omega}(R_{xx}-R_{yy}),
\\
&\displaystyle\frac{d\sigma_{zz}}{d\Omega}&=\displaystyle\frac{\alpha ^2\beta ^3}{4q^2}
\displaystyle\frac{3\tau }{4}\left [(1+\cos^2\theta )|G_M|^2+8\sin^2\theta 
\left (
\displaystyle\frac{\tau }{3}|G_Q|^2-Re(G_CG_Q^*)\right ) \right], \\
&\displaystyle\frac{d\sigma_{xz}}{d\Omega}&=-\displaystyle\frac{\alpha ^2\beta ^3}{4q^2}
3\tau ^{3/2}\sin(2\theta )Re(G_QG_M^*),
\\
&\displaystyle\frac{d\sigma_{xx}}{d\Omega}&=-\displaystyle\frac{\alpha ^2\beta ^3}{4q^2}
\displaystyle\frac{3\tau }{4}\sin^2\theta |G_M|^2, 
\end{eqnarray}

\end{itemize}

The part of the differential cross section that depends on the correlation between the longitudinal polarization 
of the electron beam and the $\rho $--meson tensor polarization can be written as follows
\be\label{eq:eq21a}
\displaystyle\frac{d\sigma_{\lambda T}}{d\Omega}=\displaystyle\frac{\alpha ^2\beta ^3}{4q^2}
6\tau ^{3/2}\sin\theta Im(G_MG_Q^*)R_{yz}.
\ee
In the experimental study of reaction (\ref{eq:eq1}), one does not measure the polarization of the outgoing particle as in the case of stable particles. Their polarization is obtained through the measurement of the angular distribution of the decay products, which allows to determine the individual elements of the spin density matrix. Therefore, the discussion of the necessary observables and of the strategy of measurements is done in the following chapter where the explicit expressions of the spin density matrix elements are given in terms of FFs.

Let us note here that, in principle, one should take into account the problem of the two--photon--exchange contribution, 
which, may become important at large momentum transfer squared, as it is suggested a few decades ago \cite{Gu73}.  Model independent properties of the two--photon--exchange 
contribution in elastic electron--deuteron scattering have been derived in Ref. \cite{Re99}. As it was shown in Ref. 
\cite{Pu61}, if the detection of the final particles does not distinguish between $\rho^- $ and $\rho^+ $--mesons, 
then the interference between one--photon and two--photon amplitudes does not contribute to the cross section of the 
reaction (\ref{eq:eq1}).

%%%%%%%%%%%%%%%%%%%%%%%%%%%%%%%%%%%%%%%%%%%%%%%%%%%%%%%%%%%
\section{Spin density matrix of $\rho$--meson}
%%%%%%%%%%%%%%%%%%%%%%%%%%%%%%%%%%%%%%%%%%%%%%%%%%%%%%%%%%%
\hspace{0.7cm}

Let us calculate the elements of the spin density matrix of the $\rho $--meson produced in the reaction (\ref{eq:eq1}). 

The convolution of the lepton $L_{\mu\nu}$ and hadron $W_{\mu\nu}$ tensors can be written as (for the case of 
unpolarized initial lepton beams)
\be
S^{un}=S_{\mu\nu}U_{\mu}U_{\nu}^*,
\label{eq:s1}
\ee
where $U_{\mu}$ is the polarization four--vector of the detected $\rho-$ meson and the $S_{\mu\nu}$ tensor can be 
represented in the following general form
\be
S_{\mu\nu}=S_1g_{\mu\nu}+S_2q_{\mu}q_{\nu}+S_3k_{1\mu}k_{1\nu}+
S_4(k_{1\mu}q_{\nu}+q_{\mu}k_{1\nu})+iS_5(k_{1\mu}q_{\nu}-q_{\mu}k_{1\nu}),
\label{eq:s2}
\ee
where the structure functions $S_i$ $(i=1-5)$ can be written in terms of three electromagnetic
FFs of the $\rho -$ meson. Their explicit form is
\ba
S_1&=&4M^2(1-\tau )q^2\bigl [\tau |G_M|^2+\sin^2\theta
|G_C+\displaystyle\frac{2}{3}\tau  G_Q|^2\bigr ], \nn\\
S_2&=&q^2\bigl [2\tau\beta \cos\theta +(1-\tau )(1+\cos^2\theta )\bigr ]|G_M|^2
+4q^2\bigl [\tau (\beta-\cos\theta)\cos\theta ReG_MG_Q^*+ \nn\\
&& +\sin^2\theta (\displaystyle\frac{\tau}{3}|G_Q|^2-ReG_CG_Q^*)\bigr ],\nn\\
S_{3}&=&4(1-\tau )q^2|G_M|^2,~
\S_4=-2q^2\bigl [(1-\tau +\tau\beta \cos\theta)|G_M|^2+
2\tau\beta \cos\theta ReG_MG_Q^*\bigr ],\nn\\
S_5&=&-4q^2\beta \cos\theta ImG_M \left (G_C-\displaystyle\frac{\tau}{3}G_Q\right )^*.
\label{eq:eq23}
\ea
The T--odd structure function $S_5$ is not zero 
here since the electromagnetic FFs of the $\rho -$ meson are complex functions in this case (TL region).

Then the elements of the spin density matrix of the $\rho -$ meson are defined as
\be\label{4}
S\rho_{mm^{\prime}}=S_{\mu\nu}U_{\mu}^{(m)}U_{\nu}^{(m^{\prime})*}, \ \
S=S_{\mu\nu} \left (-g_{\mu\nu}+\displaystyle\frac{p_{1\mu}p_{1\nu}}{M^2}\right ),
\ee
where $S=8M^2(\tau -1)q^2D$, and $U_{\mu}^{(m)}$ is the polarization
four--vector of the $\rho -$ meson with definite $(m=0, \pm 1)$ projection
on $z$ axis. In our case it is directed along the $\rho -$ meson momentum
and thus $U_{\mu}^{(m)}$ are the polarization vectors with definite helicity.

So, the elements of the spin density matrix of the $\rho -$ meson are
\ba
\rho_{++}&=&\rho_{--}=\displaystyle\frac{1}{4D}\bigl [\tau (1+\cos^2\theta )|G_M|^2+
2\sin^2\theta |G_C+\displaystyle\frac{2}{3}\tau G_Q|^2\bigr ], \nn \\
\rho_{00}&=&\displaystyle\frac{1}{2D}\bigl [\tau (1+\cos^2\theta )|G_M|^2+
\sin^2\theta |G_C-\displaystyle\frac{4}{3}\tau G_Q|^2\bigr ], \nn \\
\rho_{+-}&=&\rho_{-+}=\displaystyle\frac{\tau}{4D}\sin^2\theta |G_M|^2,\nn \\
\rho_{+0}&=&\displaystyle\frac{1}{D}\sqrt{\displaystyle\frac{\tau}{2}}\sin\theta \cos\theta
\bigl [Re G_M \left (G_C+\displaystyle\frac{2}{3}\tau G_Q \right )^*-
\left (G_C- \displaystyle\frac{\tau}{3}G_Q\right )G_M^*\bigr ], \nn \\
\rho_{-0}&=&-\rho_{+0}, ~\rho_{0+}=\rho_{+0}^*, ~ 
\rho_{0-}=\rho_{-0}^*. \nn
\label{eq:eq25} 
\ea
The spin density matrix is normalized as $Tr\rho =1$ or $\rho_{++}+
\rho_{--}+\rho_{00}=1.$ The element $\rho_{+0}$ is complex and we have
$$Re\rho_{+0}=\displaystyle\frac{1}{D}\left (\displaystyle\frac{\tau}{2}\right )^{3/2}\sin2\theta ReG_MG_Q^*, \ \
Im\rho_{+0}=\displaystyle\frac{1}{2D}\sqrt{\displaystyle\frac{\tau}{2}}\sin2\theta
ImG_M\left (G_C-\displaystyle\frac{\tau}{3}G_Q \right)^*. $$

Let us consider the case when the electron beam is longitudinally polarized. Then the convolution 
of the spin--dependent part of the lepton tensor and hadron one can be written as
\be
S(\lambda )=S_{\mu\nu}(\lambda )U_{\mu}U_{\nu}^*,
\label{eq:eqs5}
\ee
where the 
$S_{\mu\nu}(\lambda )$ tensor can be written as
\be
S_{\mu\nu}(\lambda )=Q_1\epsilon_{\mu\nu\alpha\beta}k_{1\alpha}k_{2\beta}+
Q_2(q_{\mu}a_{\nu}-q_{\nu}a_{\mu})+Q_3(q_{\mu}a_{\nu}+q_{\nu}a_{\mu}),
\label{eq:eqs6}
\ee
where $a_{\mu}=\epsilon_{\mu\alpha\beta\gamma}p_{\alpha}k_{1\beta}k_{2\gamma}$,
$p=p_1-p_2$ and the structure functions $Q_i (i=1-3)$ can be written in
terms of the $\rho -$ meson FFs as:
\ba Q_{1}&=&2i\lambda (1-\tau )q^2|G_M|^2, \nn \\
Q_{2}&=&-2i\lambda \bigl [\tau |G_M|^2-2ReG_M 
\left (G_C-\displaystyle\frac{\tau}{3}G_Q \right )^*\bigr ],\nn \\
Q_3&=&4\lambda\tau ImG_MG_Q^*. 
\label{eq:eqs7}
\ea
The T--odd structure function $Q_3$ is not zero since FFs are complex functions in the TL region.

The elements of the spin density matrix of the $\rho -$ meson that depend
on the longitudinal polarization of the electron beam can be defined as
\be\label{9}
S\rho_{mm^{\prime}}(\lambda)=
S_{\mu\nu}(\lambda )U_{\mu}^{(m)}U_{\nu}^{(m^{\prime})*}.
\ee
After some calculations we obtain
$$\rho_{++}(\lambda )=-\rho_{--}(\lambda )=\displaystyle\frac{\lambda}{2D}
\tau \cos\theta |G_M|^2, $$
$$\rho_{00}(\lambda )=\rho_{+-}(\lambda )=\rho_{-+}(\lambda )=0, $$
\be\label{10}
\rho_{+0}(\lambda )=-\displaystyle\frac{\lambda}{D}\sqrt{\displaystyle\frac{\tau}{2}}
\sin\theta \bigl [ReG_M \left (G_C+\displaystyle\frac{2}{3}\tau G_Q\right )^*-\tau G_QG_M^*\bigr ],
\ee
$$\rho_{0+}(\lambda )=\rho_{+0}^*(\lambda ), \ \
\rho_{-0}(\lambda )=\rho_{+0}(\lambda ), \ \
\rho_{0-}(\lambda )=\rho_{-0}^*(\lambda ). $$

The element $\rho_{+0}(\lambda )$ is complex quantity and its real and
imaginary parts are
$$Re\rho_{+0}(\lambda )=-\displaystyle\frac{\lambda}{D}\sqrt{\displaystyle\frac{\tau}{2}}sin\theta
ReG_M \left (G_C-\displaystyle\frac{\tau}{3}G_Q\right )^*. $$
$$Im\rho_{+0}(\lambda )=-\displaystyle\frac{\lambda}{D}\sqrt{\displaystyle\frac{\tau}{2}}
\tau \sin\theta ImG_MG_Q^*. $$

The $\rho -$ meson FFs in the TL region are complex functions.
In the case of unpolarized initial and final particles the differential cross
section depends only on the squared moduli $|G_M|^2$ and on the combination
$G = |G_C|^2+\displaystyle\frac{8}{9}\tau ^2|G_Q|^2.$ So, the measurement of the angular
distribution of the cross section allows to determine $|G_M|$ and the quantity $G$.

Let us discuss what information can be obtained by measuring the elements of
the spin density matrix of the produced $\rho -$ meson. There are three phase differences for three FFs, which we note as follows: $\alpha_1=
\alpha_M-\alpha_Q,$ $\alpha_2=\alpha_M-\alpha_C,$ and $\alpha_3=
\alpha_Q-\alpha_C,$ where $\alpha_M=ArgG_M,$ $\alpha_C=ArgG_C,$ and
$\alpha_Q=ArgG_Q$. They are related by the  condition: $\alpha_3=\alpha_2-\alpha_1$. These quantities characterize the strong interaction between final particles.

Consider the ratio of the following elements of the spin density matrix
$Re\rho_{+0}$ (when the electron beam is unpolarized) and $Im\rho_{+0}(\lambda )$
(let us remind that measurement of this element requires a longitudinally
polarized electron beam). As a result we have for this ratio
\be\label{11}
R_1=\displaystyle\frac{Re\rho_{+0}}{Im\rho_{+0}(\lambda )}=
-\displaystyle\frac{\cos\theta}{\lambda}\cot\alpha_1.
\ee
So, the measurement of this ratio gives us information about the phase
difference $\alpha_1.$ The measurement of the ratio of other
spin--density matrix elements ($Re\rho_{+0}$ and $\rho_{+-}$)
\be\label{12}
R_2=\displaystyle\frac{Re\rho_{+0}}{\rho_{+-}}=
2\sqrt{2\tau }\cot\theta \cos\alpha_1\displaystyle\frac{|G_Q|}{|G_M|}
\ee
gives us information about the quantity $|G_Q|.$ This allows to obtain the
modulus of the charge form factor, $|G_C|$, from the quantity $G$ derived by the angular distribution of the differential cross section. The
measurement of the next ratio
\be\label{13}
R_3=\displaystyle\frac{Im\rho_{+0}}{Re\rho_{+0}(\lambda )}=
-\displaystyle\frac{\cos\theta}{\lambda}\displaystyle\frac{\sin\alpha_2-r\sin\alpha_1}
{\cos\alpha_2-r\cos\alpha_1}, \ \ r=\displaystyle\frac{\tau }{3}\displaystyle\frac{|G_Q|}{|G_C|}
\ee
allows to determine the phase difference $\alpha_2.$ And at last, if we
measure the ratio of the spin--density matrix elements $\rho_{++}$ and
$\rho_{+-}$
\be\label{14}
R_4=\displaystyle\frac{\rho_{++}}{\rho_{+-}}=-\displaystyle\frac{1}{\tau\sin^2\theta}
[\tau (1+\cos^2\theta)+2\sin^2\theta
\displaystyle\frac{|G_C|^2}{|G_M|^2}(1+4r^2+4r\cos\alpha_3)]
\ee
we can obtain information about the third phase difference $\alpha_3.$

The correctness of the determination of the phase differences can be verified
by checking the relation: $\alpha_3=\alpha_2-\alpha_1$

Thus, the measurement of the spin--density matrix elements  considered above allows to obtain all 
information about the $\rho -$ meson FFs in the TL region. Remind that for the complete determination 
of FFs we need the longitudinally polarized electron beam.

%%%%%%%%%%%%%%%%%%%%%%%%%%%%%%%%%%%%%%%%%%%%%%%%%%%%%%%%%
\section{Model of $\rho$--meson form factors}
%%%%%%%%%%%%%%%%%%%%%%%%%%%%%%%%%%%%%%%%%%%%%%%%%%%%%%%%%

\hspace{0.7cm}
    
In order to predict the polarization observables for the reaction (\ref{eq:eq1}) one needs to know the behavior of the real and 
imaginary parts of all three $\rho$--meson FFs in the TL region. Unfortunately, up to now there are no data on the 
$\rho$--meson FFs, or any model for them, which works in the TL region. Therefore, we constructed a very simple model 
of the $\rho$--meson FFs, which fulfills basic known properties of these FFs, and can be extended to the TL 
region with non--zero imaginary part. This model is in a good agreement with an existing model of $\rho$--meson 
electromagnetic FFs in the SL region \cite{MF97}.

As it was shown in Ref. \cite{BH92}, the dominance of helicity--conserving amplitudes in gauge theory implies universal 
ratios for the charge, magnetic, and quadrupole FFs of spin--one bound states:
\be
G_C(Q^2):G_M(Q^2):G_Q(Q^2)=\left(1-\displaystyle\frac{2}{3}\eta\right):2:-1 \label{asymratio}
\ee
where $\eta =Q^2/4m^2$ ($m$ is the mass of the spin--one particle). These ratios hold at large SL or TL momentum transfer 
in the case of composite systems such as the $\rho$--meson or deuteron in QCD with corrections of order 
$\Lambda_{QCD}/Q$ and $\Lambda_{QCD}/M_{\rho,d}$, where $\Lambda_{QCD}$ is the QCD scale. 

The dimensional counting rules \cite{Ma73} of pQCD for exclusive two--body scattering processes at large $s$, with 
$t/s$ fixed, predict the following asymptotic behavior of the differential cross section
\be
\displaystyle\frac{d\sigma}{dt}\Big |_{t\rightarrow \infty} \sim\displaystyle\frac{1}{t^{n-2}}f(t/s), \label{asymp}
\ee
where $n$ is the total number of incoming and outgoing fields ($n=6$ in the case of $e^+e^-\rightarrow \rho^+\rho^-$ 
or $e^-\rho\rightarrow e^-\rho$ reactions).

These asymptotic conditions imply the following asymptotic behavior of the $\rho$--meson FFs $G_C(t), G_M(t), G_Q(t)$ (in our case 
$t=q^2$)
\be
|G_C(t)|\Big |_{t\rightarrow \infty} \sim t^{-1} \; , \; |G_M(t)|\Big |_{t\rightarrow \infty} \sim t^{-3/2} \; , 
\; |G_Q(t)|\Big |_{t\rightarrow \infty} \sim t^{-2}. \label{asymFF}
\ee
The normalization of FFs, $G_C(t), G_M(t), G_Q(t)$ at  $t=0$, is consistent with the calculation \cite{MF97}
\be
G_C(0)=1 \; , \; G_M(0)=\mu_{\rho} =2.14 \; , \; G_Q(0)=-m_{\rho}^2Q_{\rho}=-0.79.
\ee
The most simple parametrization of the charge $\rho$--meson electromagnetic FFs, is a monopole parametrization. However, it is not possible 
to obtain a node, as predicted  by impulse approximation models \cite{MF97} using a simple monopole/dipole parametrization (monopole/dipole functions are always different 
from zero).  We used the following parametrization
\be
G_C(t)=\displaystyle\frac{(A+Bt)G_C(0)}{1-\displaystyle\frac{t}{m_{C}^2}}, ~
G_M(t)=\displaystyle\frac{G_M(0)}{\left (1-\displaystyle\frac{t}{m_{M}^2}\right )^2}, ~
G_Q(t)=\displaystyle\frac{G_Q(0)}{\left (1-\displaystyle\frac{t}{m_{Q}^2}\right )^2},
\label{eq:eqffsl}
\ee
where $A$, $B$, $m_C$, $m_M$ and $m_Q$ are free parameters, which were fitted in order to reproduce the  $\rho$ meson FFs in SL region given by \cite{MF97}. The values of parameters $A$ and $B$ were fixed by the normalization and the position of the node $t_0\simeq -3$ GeV$^2$
\be
G_C(0)=A=1,~A+Bt_0 =0~\to~ \\
B=-\displaystyle\frac{A}{t_0}.
\ee
For simplicity, we take a dipole form for  $G_M(t)$ and $G_Q(t)$. Therefore, the behavior for $G_M(t)$ does not follow precisely the asymptotic limits as described above (\ref{asymFF}). This simple parametrization is chosen in order to calculate observables in a restricted energy range, and the values of the parameters (Table 1) are fitted on the model predictions from SL region \cite{MF97}. 

\begin{table}
\begin{tabular}{|c|c|c|c|c|}
\hline
$A$ & $B$   & $m_C$ [GeV]& $m_{M}$ [GeV]&$m_{Q}$ [GeV] \\
\hline
$1.$ & $0.33$ & $1.34$ &  $1.42$ & $1.51$ \\
\hline
\end{tabular}
\caption{ Parameters of the model for $\rho$--meson electromagnetic FFs.}
\label{tab1}
\end{table}

The extension of the model to TL region was made by introducing widths for the  particles carrying the interaction. This leads to the following parametrization, which contains an imaginary part:
\be
G_C(t)=\displaystyle\frac{(A+Bt)m_C^4}{(m_C^2-t -im_C\Gamma_C)^2},~  
G_M(t)=\displaystyle\frac{G_M(0)m_M^4}{(m_M^2-t -im_M\Gamma_M)^2},~
G_Q(t)=\displaystyle\frac{G_Q(0)m_Q^4}{(m_Q^2-t -im_Q\Gamma_Q)^2}. 
\label{eq:eqtl}
\ee
%%%%%%%%%%%%%%%%%%%%%%%%%%%%
\section{Results}
%%%%%%%%%%%%%%%%%%%%%%%%%%%%%%%%%%%%%%%%%%%%
The $\rho$ meson FFs have been calculated in both SL and TL regions for two different values of the widths $\Gamma_C$, $\Gamma_{M}$, and $\Gamma_{Q}$, taken as 1\% and  10\% of the corresponding masses. In TL region, FFs are complex, and present, in correspondance of the value of the parameters, a resonant-like behavior which enhances the three FFs. Therefore, the absolute value of the cross section is extremely sensitive to a small variation of the parameters, in this region. 

\begin{figure}
\centering
\includegraphics[width=16cm]{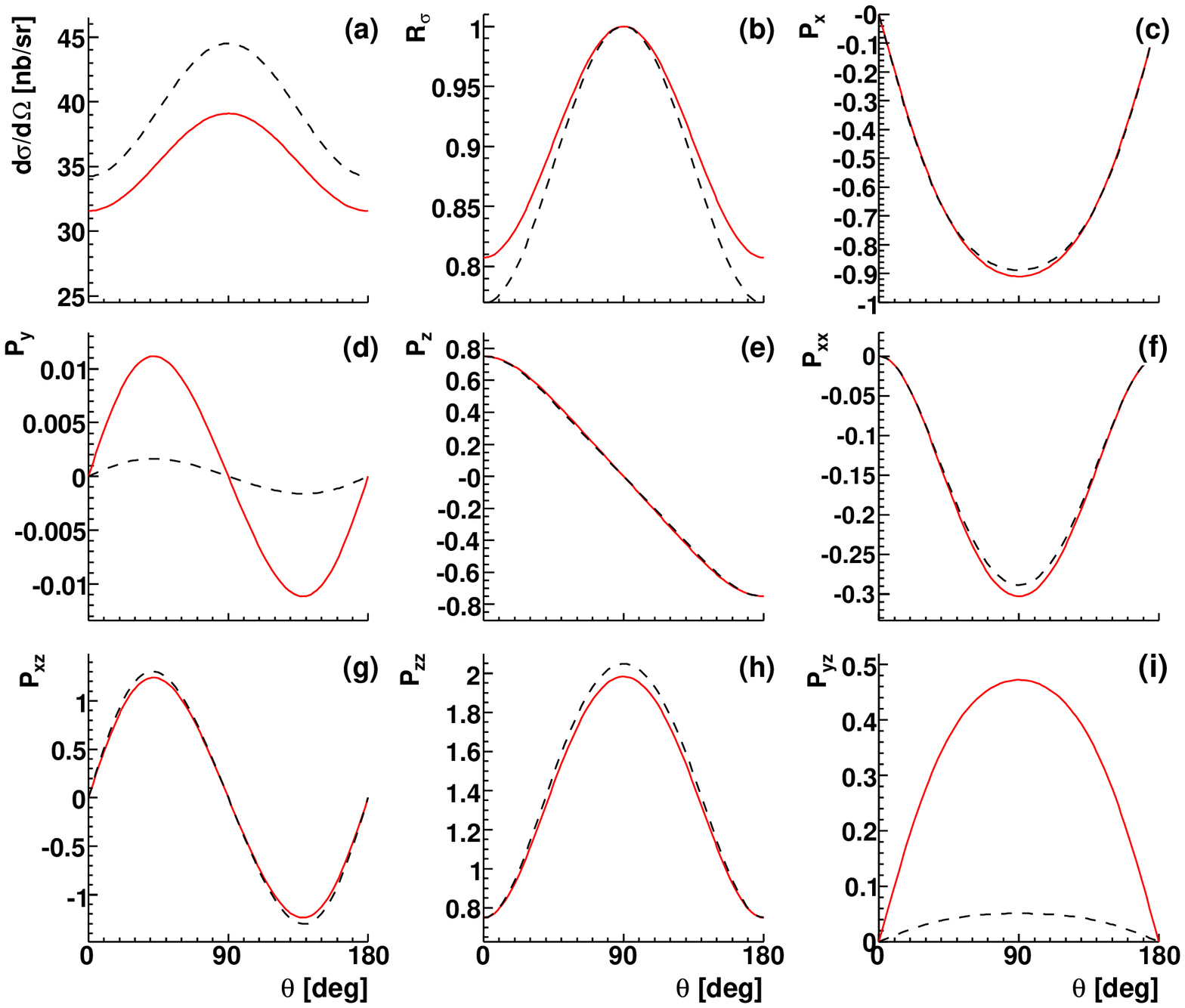}
\caption{(Color online) Angular dependence of the  differential cross section (a); of the angular asymmetry $R_{\sigma}$ Eq. (\protect\ref{eq:eq20}) (b); of the  spin polarization observables: $P_x$ (c), $P_y$ (d), $P_z$ (e), $P_{xx}$:(f), $P_{yy}$ (g), $P_{zz}$ (h). The full line corresponds to a 10\% width , the  dashed line to 1\% width.}
\label{fig:rhopolar}
\end{figure}

The numerical results are shown for $q^2$ =3 GeV$^2$. The differential cross section is shown in Fig. \ref{fig:rhopolar}a and the angular asymmetry in Fig. \ref{fig:rhopolar}b. The shape of these distributions is a signature of the one photon mechanism. The absolute value change by 30\% in case of 10\% width (dashed line), compared to 1\% width (solid line). But the comparison at slightly different $q^2$ values can differ by order of magnitudes, due to the presence of the resonances in the FFs parametrization. The different polarization observables are shown in Fig. \ref{fig:rhopolar}c-i. The behavior of $P_x$, $P_z$, $P_{xx}$, and  $P_{xz}$ is almost independent on the width, as these observables are determined either by the real part or by the modulus of FFs.

\begin{figure}
\centering
\includegraphics[width=16cm]{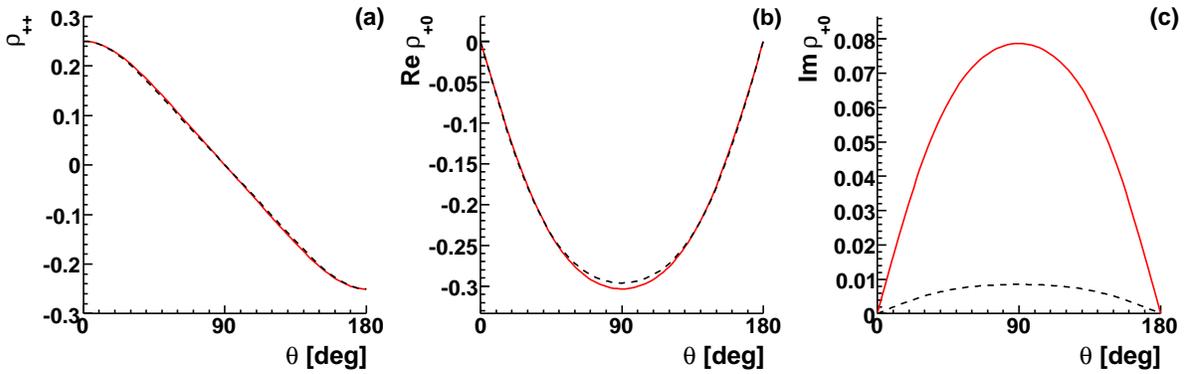}
\caption{(Color online) Angular dependence of the  elements of the $\rho$ density matrix:
$\rho_{++}(\lambda)$ (a), Re$\rho_{+0}(\lambda)$ (b), Im$\rho_{+0}(\lambda)$  (c). The full line corresponds to a 10\% width , the  dashed line to 1\% width. }
\label{fig:densityRho}
\end{figure}

The elements of the (polarized) spin density matrix of the $\rho$ meson are plotted in Fig. \ref{fig:densityRho}, at $q^2$ =3 GeV$^2$: $\rho_{++}(\lambda)$ is shown in  Fig. \ref{fig:densityRho}a, the real and imaginary parts of  $\rho_{+0}(\lambda)$ in Figs. \ref{fig:densityRho}b and \ref{fig:densityRho}c ,  respectively. The magnitude of these last two terms is quite sensitive to the value of the width, whereas $\rho_{++}(\lambda)$ is not, because it depends on  $|G_M|^2$.

All these observables are very sensitive to different combinations of FFs, therefore their measurement will be especially discriminative toward FFs models.
%%%%%%%%%%%%%%%%%%%%%%%%%%%%%
\section{Conclusion}
%%%%%%%%%%%%%%%%%%%%%%%%%%%%%
Using the parametrization of the electromagnetic current for $\gamma^*\rho\rho $ vertex in terms
of three complex FFs,  we investigated the polarization phenomena in the reaction (\ref{eq:eq1}). We 
calculated for these reactions the differential (and total) cross sections and various 
polarization observables as functions of the corresponding set of FFs. The spin--density matrix 
elements of the produced $\rho $ meson for the reaction (\ref{eq:eq1}) have been also calculated.

We constructed a simple model for the $\rho $ meson FFs and fitted free parameters of this model 
to the predicted values of these FFs which were calculated (for the SL region) both in covariant 
and light--front frameworks with constituent quarks \cite{MF97}. Then FFs of our model were 
analytically continued to the TL region. Using this model we estimated the differential cross 
section and various polarization observables which were found to be sizeable.

The reaction (\ref{eq:eq1}) has not yet been detected in the existing experiments on colliding electron--positron beams in the $q^2$ region between 1 and 4 GeV$^2$.  We showed that, in frame of VMD models, the absolute value is very sensitive to the presence, the position and the width of resonances in this momentum range.  So, the experimental investigation of this reaction 
will give very useful information for the understanding of the electromagnetic properties of 
the $\rho $ meson and constitute a good testing for models of the $\rho $ meson FFs.

\section{Aknowledgments}
The work of one of us (G.I.G) is supported in part by the INTAS grant, under Ref. Nr. 05-1000008-8328.
The Slovak Grant Agency for Sciences VEGA is acknowledged by C.A., for support under Grant N. 2/4099/26.

\end{document}